\documentclass[12pt]{article}
\usepackage{array,multicol}
\usepackage{afterpage,float,flafter}
\usepackage[dvips]{epsfig,rotating}
\usepackage{pifont,fancybox}
\usepackage{latexsym}

\newcommand{\be}{\begin{equation}}
\newcommand{\ee}{\end{equation}}

\newcommand{\bea}{\begin{eqnarray}}
\newcommand{\eea}{\end{eqnarray}}
\newcommand{\eq}[1]{eq.~(\ref{#1})}

\newcommand{\gsim}{\ \rlap{\raise 2pt\hbox{$>$}}{\lower 2pt \hbox{$\sim$}}\ }
\newcommand{\lsim}{\ \rlap{\raise 2pt\hbox{$<$}}{\lower 2pt \hbox{$\sim$}}\ }

\newcommand{\matr}{\left( \begin{array}}
\newcommand{\ematr}{\end{array} \right)}

\def\bq{\begin{quote}}
\def\eq{\end{quote}}
\def\ben{\begin{enumerate}}
\def\een{\end{enumerate}}

\def\ie{{\it i.e.}}

\def\sin{{\rm sin}}

\def\Dslash{\not{\hbox{\kern-4pt $D$}}}
\def\dslash{\kern-4pt \not{\hbox{\kern-2pt $\partial$}}}
\def\pslash{\not{\hbox{\kern-2pt p}}}

\makeatletter
\@addtoreset{equation}{section}
\makeatother

\title{
\vspace*{1.0cm}
{$CPT$ violating neutrinos
in the light\\ of KamLAND and Super-K}
\vspace*{0.8cm}
\author{\large\textbf
{G.~Barenboim$^a$, L.~Borissov$^b$, J.~Lykken$^{a,c}$}\\ 
\\
$^a$\normalsize\emph{Fermi National Accelerator Laboratory,
P.O. Box 500, Batavia, IL 60510, USA }\\
$^b$\normalsize\emph{Columbia University, New York, NY, 10027, USA}\\
$^c$\normalsize\emph{Enrico Fermi Institute, Univ. of Chicago, 5640
S. Ellis Ave., Chicago, IL 60637, USA}\\ }
}

\begin{document}
\maketitle

\vspace*{0.6cm}

\begin{abstract}
We show that a hierarchical $CPT$ violating
neutrino spectrum can simultaneously accommodate all published
neutrino data, including the LSND result, with a total $\chi^2$
which is almost identical to the $CPT$ conserving best fit.
In our scenario
the oscillation signal seen by the KamLAND experiment in antineutrinos
is independent of the LMA solar oscillation signal seen in neutrinos.
A larger antineutrino mass splitting accounts for the LSND
signal and also contributes to atmospheric oscillations. Because of
this feature, there is some tension between our model and certain Super-K
atmospheric results. Thus, if LSND did not exist, our model would
survive only at the 99\% confidence level; with LSND, our model is
(essentially) statistically equivalent to the $CPT$ conserving best fit.

\end{abstract}

\thispagestyle{empty}
\newpage

\section{Introduction}

$CPT$ violating neutrino masses allow the possibility 
\cite{Murayama} - \cite{Nos3}
of reconciling the LSND \cite{lsnd}, atmospheric \cite{atm_sk}, 
and solar oscillation \cite{solar, SNO_solar}
data without resorting to sterile neutrinos. As argued in
\cite{Nos1}, 
there are good reasons to imagine that $CPT$ violating
dynamics couples directly to the neutrino sector, but not
to other Standard Model degrees of freedom. 
An explicit $CPT$ violating model of this type was presented
in \cite{Nos3}.

KamLAND \cite{kamland}, a medium baseline reactor antineutrino disappearance
experiment, is sensitive to antineutrino mass-squared splittings 
in the $10^{-4}$ eV$^2$ range characteristic of the large mixing
angle (LMA) solar neutrino scenario. The KamLAND collaboration has
recently reported \cite{kamland2}
an electron antineutrino survival probability
which is significantly less than one:
\be
P(\overline{\nu}_e \rightarrow \overline{\nu}_e) = 
0.611 \pm 0.085 \pm 0.041 \; .
\label{kamdata}
\ee

If the neutrino mass spectrum conserves $CPT$, then this result
is consistent with the LMA interpretation of solar neutrino
oscillations. If the neutrino mass spectrum violates $CPT$,
however, the KamLAND result provides no information about
solar oscillations, but rather constrains the splittings
in the antineutrino spectrum.

In this paper we show that a hierarchical $CPT$ violating
neutrino spectrum can simultaneously accommodate the
oscillation data from LSND, atmospheric, solar and KamLAND,
as well as the nonobservation of antineutrino
disappearance in short baseline reactor experiments. 
Comparing our model to the global neutrino data set we obtain
a total $\chi^2 = 201$ for 228 degrees of freedom. This can
be compared to the usual $CPT$ conserving best fit, which
has a total $\chi^2 = 199$ for 232 degrees of freedom
\cite{Gonzalez-Garcia:2003jq}.

While the total $\chi^2$ are about the same, there are striking
differences when we make a more detailed comparison to the data.
The $CPT$ conserving best fit parameter set (two mass differences
and 3 mixing angles) matches remarkably well to the global data
set, with the glaring exception of the LSND result. This single
data point contributes a $\Delta\chi^2$ of about +12, leading to
widespread speculation that the LSND result is simply wrong, and
will be disconfirmed by MiniBooNE.

As we will see, the hypothesis of a $CPT$ violating
neutrino sector (four mass differences and 6 mixing angles) leads
to a completely different conclusion. In this framework our
(non-optimized) parameter choices give an excellent fit to
LSND. The largest single discrepancy with the global data set
instead occurs with the atmospheric Super-K multi-GeV muons
and thru-going muons. In either case the $\Delta\chi^2$ is
no worse than +4. Thus instead of casting a skeptical eye upon
LSND one is led to speculate that the forthcoming re-analysis
of the Super-K atmospheric data will improve this discrepancy.
Indeed the new fluxes announced for the Super-K re-analysis
definitely work in this direction.

We also make a dramatic prediction for the observation of atmospheric
neutrinos using the MINOS detector \cite{minoscosmic}.
Because the MINOS detector discriminates
positive and negative charge, this experiment can
disentangle the neutrino and antineutrino components
of atmospheric oscillations in a straightforward way.
As the mass differences in the atmospheric sectors
differ by orders of magnitude in our scenario,
MINOS will be able to tell them apart easily.

\section{The spectrum}

To analyze all the possible $CPT$ violating spectra is not an easy job.
With four mass differences and six mixing angles (not taking into
account the two $CP$ violating phases which participate in oscillations)
a complete scan of the whole parameter space is impractical. 
However, thanks to
the available experimental data, it is possible to reduce the
allowed regions to two sets of well-differentiated spectra with
(quasi) orthogonal experimental signatures.

The easiest way to make contact with the experimental results 
is in terms of the neutrino survival and transition probabilities,
which are given by
\bea
P(\nu_\alpha \rightarrow \nu_\beta) = \delta_{\alpha \beta} -
 4 \sum_{i>j=1 }^3 U_{\alpha i} U_{\beta i} U_{\alpha j} 
U_{\beta j}
\;\sin^2 \left[ \frac{\Delta m_{ij}^2 L}{4 E} \right]
\label{pro}
\eea
for neutrinos and
\bea
P(\overline{\nu}_\alpha \rightarrow \overline{\nu}_\beta) = 
\delta_{\alpha \beta} -
 4 \sum_{i>j=1 }^3 \overline{U}_{\alpha i} \overline{U}_{\beta i} 
\overline{U}_{\alpha j} \overline{U}_{\beta j}
\;\sin^2 \left[ \frac{\Delta \overline{m}_{ij}^2 L}{4 E} \right]
\label{apro}
\eea
for antineutrinos.
Here the matrix $U=\left\{ U_{\alpha i}\right\}$  
($\overline{U}=\left\{ \overline{U}_{\alpha i}\right\}$)
describes the weak
interaction neutrino (antineutrino) states, $\nu_\alpha$, in terms of the 
neutrino (antineutrino) mass eigenstates,
$\nu_i$. That is,
\bea
\nu_\alpha = \sum_i U_{\alpha i} \nu_i \;\;\;\;\;\; \mbox{and} \;\;\;\;\;\;
\overline{\nu}_\alpha = \sum_i \overline{U}_{\alpha i} \overline{\nu}_i
\eea
where we have ignored the possible $CP$ violation phases in both matrices and 
took them to be real. The matrices can be
parametrized as follows:
\bea 
U=\pmatrix{c_{12} c_{13} & s_{12}c_{13} & s_{13} \cr
       -s_{12}c_{23} - c_{12}s_{23}s_{13} & c_{12}c_{23}- s_{12}s_{23}s_{13} & s_{23}c_{13} 
\cr
        s_{12}s_{23} - c_{12}c_{23}s_{13} & - c_{12}s_{23} - s_{12}c_{23}s_{13} 
&c_{23}c_{13}}
\eea
and similarly for $\overline{U}$.
In Eq.~(\ref{pro}) $L$ denotes the neutrino flight path, \ie\ 
the distance 
between the neutrino source and the detector, and
$E$ is the energy of the neutrino in the laboratory 
system. 

Regarding the mass spectrum of the three neutrinos we 
assume that it is hierarchical and thus
characterized by two different  squared masses
\[ \Delta m_{12}^2 = m^2_2 
-m^2_1\quad\mbox{and } \Delta m_{13}^2 = m^2_3 -m^2_1 \]
whose numerical values are rather different, \ie\ $
\Delta m_{13}^2 \gg  \Delta m_{12}^2$  and similarly for the 
antineutrinos. Having said that, it becomes apparent that
the larger mass-squared difference in the neutrino sector
will be related to the atmospheric neutrino signal observed
by Super-Kamiokande, while the smaller one will drive the
solar neutrino oscillations. In the antineutrino sector,
the largest mass difference will provide an explanation
to the signal observed in LSND, while the smaller one is
the one which might have been (mis)identified by KamLAND as a confirmation
of LMA.

\begin{figure}[ht]
\vspace{1.0cm}
\centering
\epsfig{file=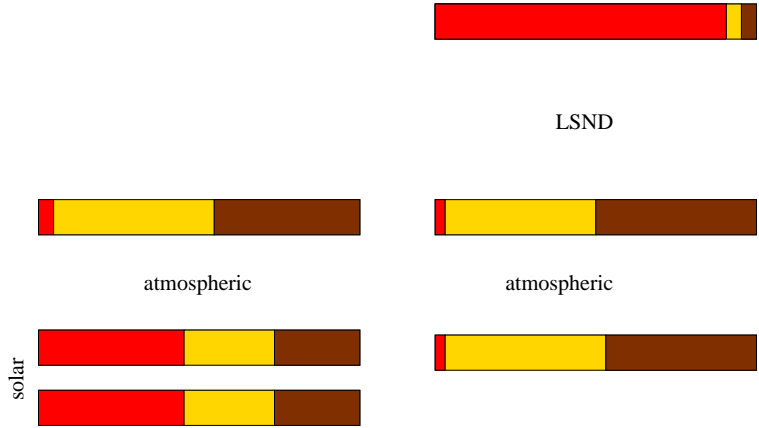,width=10cm}
\caption{\it Possible neutrino mass spectrum with almost all the 
electron content in the heavy state.  Although the figure shows an 
example of large mixing, our approach is agnostic about the mixing matrix.
The flavor content is distributed as follows: electron flavor (red),
muon flavor (brown) and tau flavor (yellow)}
\label{neutrino-spectrum}
\end{figure}

The key ingredient to sort out the antineutrino spectra are reactor 
experiments. Their results indicate \cite{chooz, bugey}
that electron antineutrinos produced
in reactors remain electron antineutrinos on short baselines. As the 
distance traveled by our antineutrinos is small we can forget about
the smallest mass difference and average the other two, thus
the survival
probability can be expressed as
\bea
P(\overline{\nu}_e \rightarrow \overline{\nu}_e) = 
1 - 2 \overline{U}_{e3}^2 (1- \overline{U}_{e3}^2) \; .
\eea
It is clear that there are two possible ways to achieve a
survival probability close to one, \ie\ $\overline{U}_{e3}$ can
be almost one or almost zero. Physically this means that we
can choose between having almost all the antielectron flavor
in the heavy state (or in the furthest away state) or just
leave in this state almost no antielectron flavor. 
The first possibility (which is depicted in Fig. 1) 
is the one we explored in our previous works. This spectrum predicts 
for KamLAND a survival probability consistent with one.
Since this is strongly disfavored by the KamLAND result
(\ref{kamdata}), we instead pursue the second possibility,
which is represented by the spectrum shown in Figure 2.

\begin{figure}[htb]
\vspace{1.0cm}
\centering
\epsfig{file=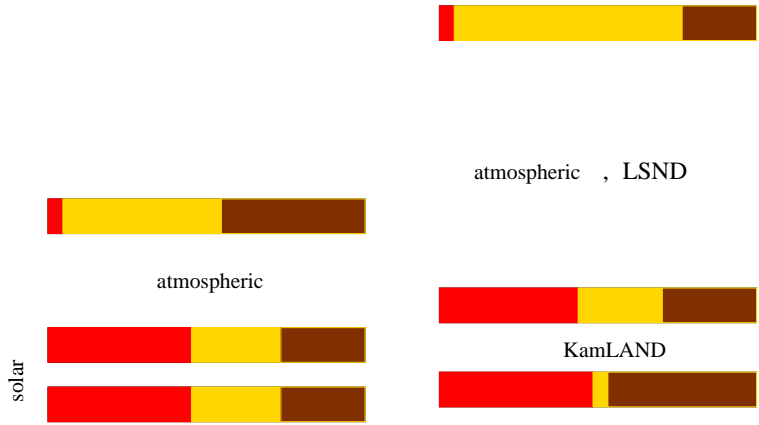,width=10cm}
\caption{\it Possible neutrino mass spectrum with almost no  
electron content in the heavy state.  Although the figure shows an 
explict mixing pattern, there is a whole family of mixing
matrices that can do an equally good job.
The flavor content is distributed as follows: electron flavor (red),
muon flavor (brown) and tau flavor (yellow)}
\label{neutrino2-spectrum}
\end{figure}

This second family of spectra is characterized by a
strong violation of $CPT$ in the mass differences
but a much slighter effect in the mixing matrix. This is
seen in Fig.~2 where  the flavor distribution in the neutrino and
antineutrino spectra is rather similar. The most
distinctive feature of this family of solutions is its  $\theta_{23}$,
which lives far away from maximal mixing, 
or in other words which has  a large component
of antitau neutrino in the heavy state.
The small antimuon neutrino component in the heavy state
is not bounded by the non observation of
muon neutrino disappearance over short
baselines in the CDHS experiment\cite{dydak}, as the antineutrino
component in this experiment was minimal.

KamLAND could have observed an oscillation signal driven
by the smaller antineutrino mass splitting
and interpreted it as LMA oscillations. To explicitly see how this
might have
happened, we will choose two sample points in our parameter space and
calculate the transition probabilities for it. Let us emphasize 
that we have not performed a chi-squared fit and therefore the points
we are selecting (by eye and not by chi) are not optimized to give
the best fit to the existing data. Instead, they must be regarded 
as two among the many equally good sons in this family of solutions.  

The point we have chosen has $\overline{\theta}_{13}=.08\;$,
$\overline{\theta}_{23}=.5\;$, $\overline{\theta}_{12}=.6\;$, 
$\Delta \overline{m}_{12}^2 = 5 \cdot 10^{-4}$ eV$^2$ and 
$\Delta \overline{m}_{13}^2 = \cal{O}$(1) eV$^2$.
Since we are dealing with an antineutrino signal, we
do not need to identify either the flavor distribution
or the mass eigenstates of the neutrino sector. We will do
it later, when showing the zenith angle dependence this model
predicts for Super-Kamiokande atmospheric neutrinos.

The survival probability measured by KamLAND is given by
\bea
P_{\mbox{\tiny{KamLAND}}} &= &1 -4 \overline{U}_{e3}^2  (1-
\overline{U}_{e3}^2) \,\sin^2 \left[ \frac{\Delta \overline{m}_{13}^2  L}{4 E} 
\right]
 -4  \overline{U}_{e1}^2  \overline{U}_{e2}^2
\,\sin^2 \left[ \frac{\Delta \overline{m}_{12}^2  L}{4 E} \right] \; ,
\eea
where the second term (proportional to $\overline{U}_{e3}^2$) is negligible.
Plugging our numbers in, it is straightforward to see that
$P_{\mbox{\tiny{KamLAND}}} \approx .6 $ regardless of whether
the mass difference that drives the solar neutrino oscillations
belongs to the LMA region.

By the same token, we can calculate the probability associated with
the LSND signal. It is given by
\bea
P_{\mbox{\tiny{LSND}}}= 4 \overline{U}_{\mu3}^2 \overline{U}_{e3}^2  
\,\sin^2 \left[ \frac{\Delta \overline{m}_{13}^2 L}{4 E} \right] \; ,
\eea
where we have neglected terms proportional to $\Delta \overline{m}_{12}^2 $
which are irrelevant for such small distances. As the reader can easily
verify, we predict a $P_{\mbox{\tiny{LSND}}} \simeq .0022$ in excellent
agreement with the LSND final analysis:
\bea
P_{\mbox{\tiny{LSND-final}}}= 0.00264 \pm .00081 \;.
\eea

The only piece of experimental evidence involving antineutrinos
which remains to be checked is the signal found for Super-Kamiokande
atmospheric neutrinos. As we are introducing an antineutrino mass
difference roughly two orders of magnitude larger than the Super-K best
fit point (for an analysis with two generations and conserving
$CPT$), there is cause for concern.
In fact we pass this test as successfully as we did the others.
To see this, we have first to state the parameters in
the neutrino sector. Once more they have been chosen almost randomly
from the different analyses available in the literature and are
given by $ \theta_{13}=.08\;$,
$ \theta_{23}=.78\;$, $ \theta_{12}=.52\;$, 
$\Delta  m_{12}^2 = 1 \cdot 10^{-4}$ eV$^2$ and 
$\Delta  m_{13}^2 = 2.8 \cdot 10^{-3}$ eV$^2$. 
We stress that although we have chosen a
point in the LMA region, the particular election of both
$\Delta  m_{12}^2 $ and  $ \theta_{12}$ does not affect the 
quality of the agreement with the data.

With these parameters we have calculated the zenith angle dependence of the 
ratio (observed/expected in the no oscillation case) for muon and
electron neutrinos for the sub-GeV and multi-GeV energy ranges 
(remember that since Super-K is a water Cherenkov
detector it does not distinguish neutrinos from antineutrinos and
washes out any possible difference between the conjugated 
channels). The results are shown in Fig. 3 where we have also included
the experimental data for the sake of comparison.
As we have closely followed the spirit of the 
calculation in  \cite{amol, Nos2},
we refer the reader to this article for details and skip the technicalities.
We worked in a complete three generation
framework and  included matter effects.

\begin{figure}[htb]
\vspace{1.0cm}
\centering
\epsfig{file=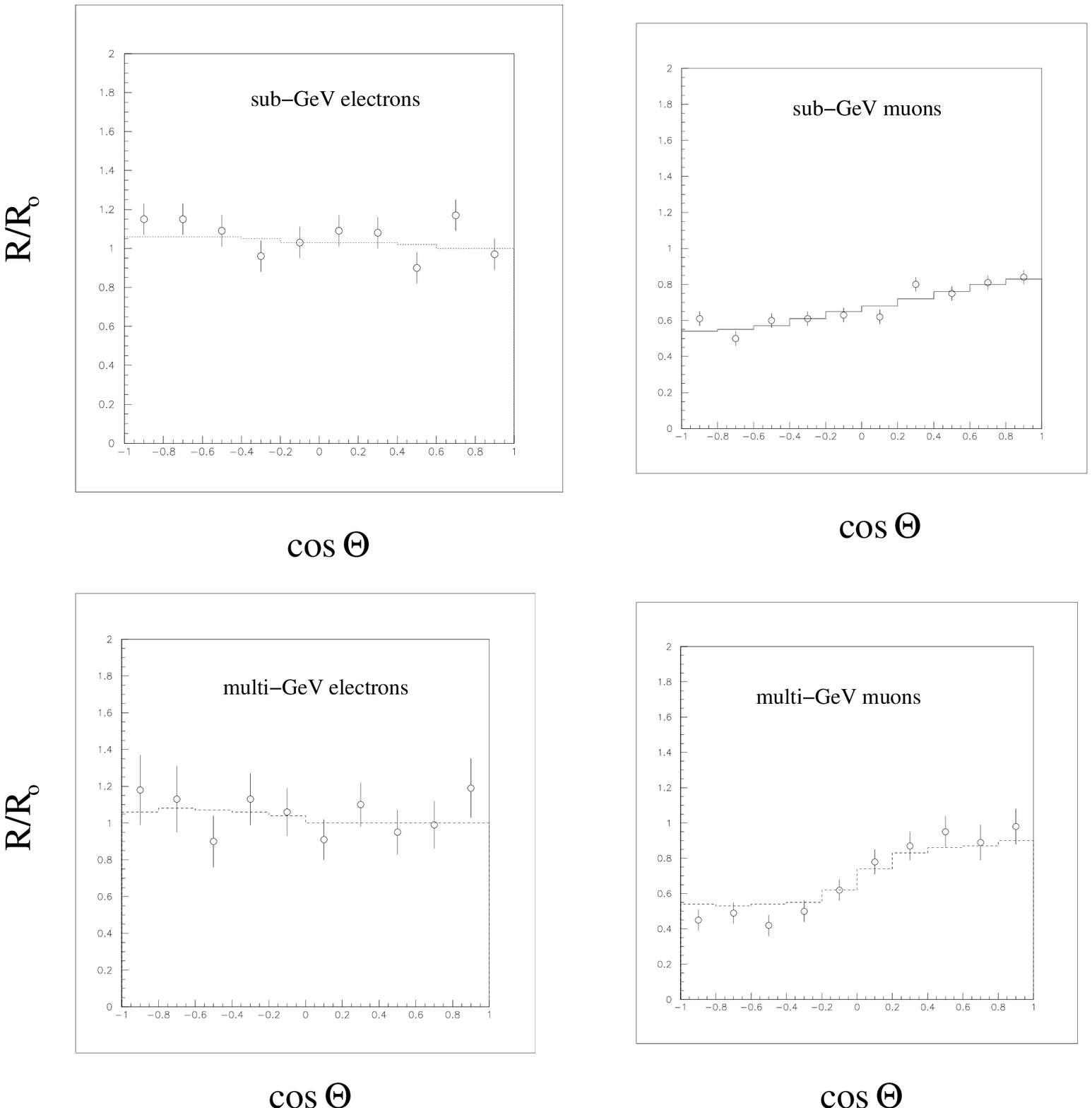,width=12cm}
\caption{\it SK zenith angle distributions normalized to no-oscillations
expectations, for our first $CPT$ violating example.
Circles with error bars correspond to SK data.}
\end{figure}

\begin{figure}[htb]
\vspace{1.0cm}
\centering
\epsfig{file=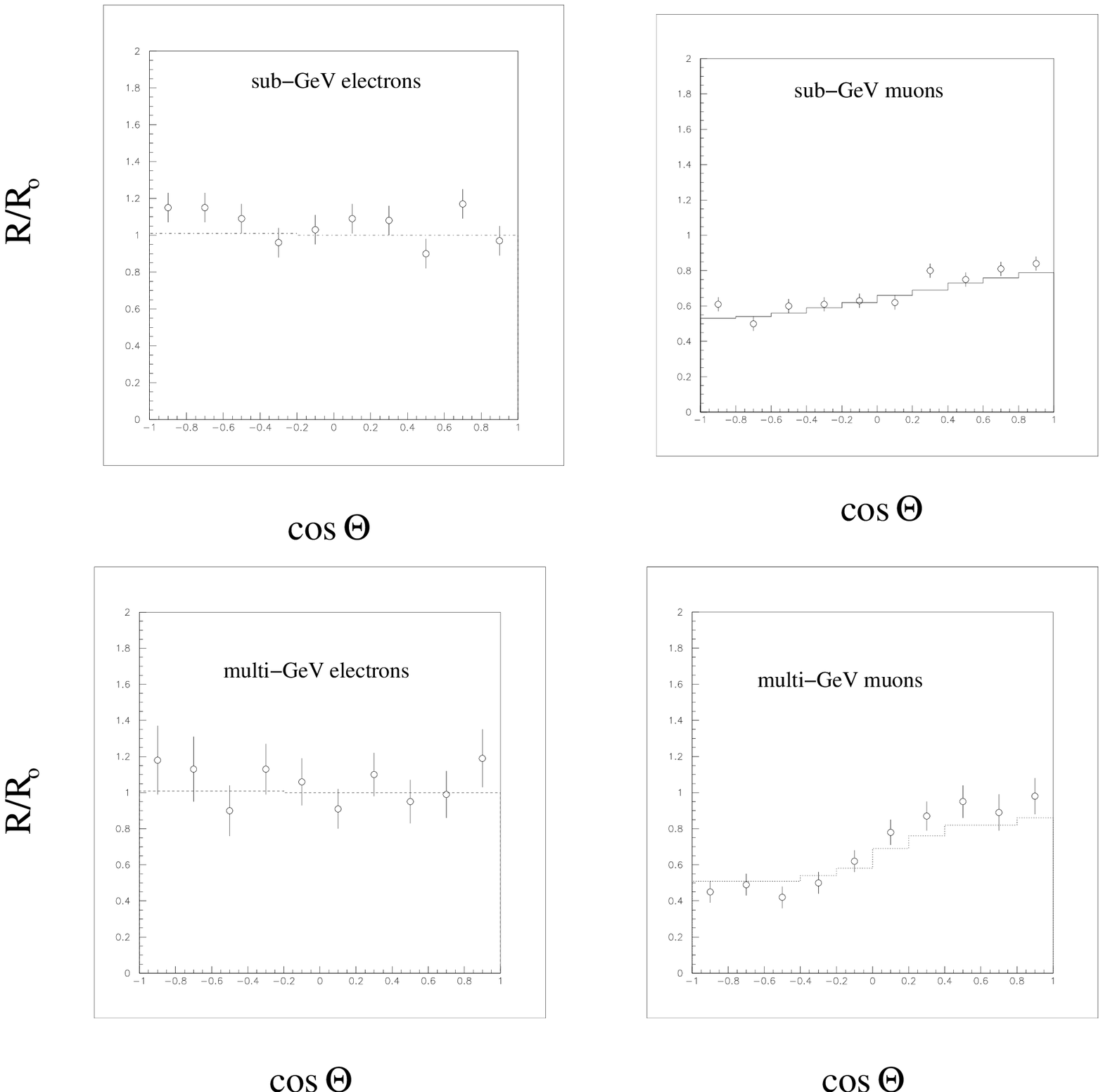,width=12cm}
\caption{\it SK zenith angle distributions normalized to no-oscillations
expectations, for our second $CPT$ violating example.
Circles with error bars correspond to SK data.}
\end{figure}

In Fig.~4 we show the comparison to Super-K for our second
example point. For this point we have chosen
$\overline{\theta}_{13}=.08\;$,
$\overline{\theta}_{23}=.5\;$, $\overline{\theta}_{12}=.785\;$, 
$\Delta \overline{m}_{12}^2 = 7 \cdot 10^{-5}$ eV$^2$ and 
$\Delta \overline{m}_{13}^2 = \cal{O}$(1) eV$^2$.
Note that this point is consistent with the best-fit point
of KamLAND \cite{kamland2}.

\section{Comparison with other analyses}

It is clear from Figs.~3 and 4 that our $CPT$ violating spectrum
does pretty well, with a moderate discrepancy apparent only for
the multi-GeV muons.
In order to understand the results it is important to remember that
due to production and cross section effects Super-K is dominated by neutrinos,
with antineutrinos a minor (but not negligible) contribution. 
This is only part of the story, however, since an
analysis done by the Super-K collaboration
allowing for $CPT$ violation did not allow (at 99\% C.L.) a 
mass difference in the antineutrino sector as drastically 
different as the one we are proposing. 
The difference is due to the fact that the Super-K
analysis was done in a two generation context and, most importantly,
forcing the two mixing angles to be maximal.
This latter fact indeed maximizes 
the antineutrino contribution and compels the antineutrino mass
difference to take the closest possible value to the neutrino
one. By the same token, if we want a large antineutrino mass
difference, we expect to improve the fit to Super-K data by
combining this with a non-maximal mixing angle, which
suppresses the antineutrino contribution to the Super-K oscillation
signal.

The two vs three generation analysis has also an impact, as is
seen by inspecting the transition probability for muon antineutrinos
into tau antineutrinos, which is given by,
\bea
P(\overline{\nu}_\mu \rightarrow \overline{\nu}_\tau) =
&= & 4 \overline{U}_{\mu 3}^2 \overline{U}_{\tau 3}^2 
\sin^2 \left[ \frac{\Delta \overline{m}^2_{23} L}{4 E} \right] 
- 4 \overline{U}_{\mu 2} \overline{U}_{\tau 2} 
\overline{U}_{\mu 1} \overline{U}_{\tau 1} 
\sin^2 \left[ \frac{\Delta \overline{m}^2_{12}  L}{4 E} \right]
\eea
From this formula it becomes apparent that for
neutrinos coming from above only the largest mass difference contributes.
However, for those neutrinos which have travelled through sizeable
portions of the Earth and have covered distances of the order of
10$^4$ km, the second mass difference also plays a role. This 
contribution (which does affect the final result, especially
for sub-GeV neutrinos) is neglected if only one mass difference is taken
into account. 

Our analysis agrees with the spirit of the findings in
Ref~\cite{solveig} where a two generation approximation that 
didn't include matter effects was used. 
Also a simplified analysis based only on the up/down asymmetry in the
number of multi-GeV events (in the $CPT$ violating case) is available in the
literature \cite{strumia}, which used an older Super-K data set.
If one uses (as we do) the result
from the full 1490 day of SK-I data, \ie \
$ A_\mu= -.288 \pm .030 $ \cite{Todd} the $CPT$ violating case (which gives
for the sample points we have being using  
$ A_\mu = -.27$) turns out to be as acceptable as the $CPT$ conserving one 
(with $ A_\mu= -.33 $ for maximal mixing but where smaller values
can be obtained by departures of maximal mixing).  
In all the cases
the electron neutrino asymmetry is consistent (within experimental
errors) with zero.

In order to get a quantitative estimate of how well
our $CPT$ violating spectrum fits the global data set {\it not
including} LSND, we have compared
(for both of our sample points) the total $\chi^2$ to the
minimum total $\chi^2$ which is obtained for different values of
the 10 input parameters. 
Fixing $\theta_{13}=\overline{\theta}_{13}=.08$ our (admittedly crude)
program finds that the minimum total $\chi^2$ occures for 
$\theta_{23}=\overline{\theta}_{23}=.8$,
$\theta_{12}=\overline{\theta}_{12}= .6$, 
$\Delta m_{12}^2 = \Delta\overline{m}_{12}^2=6\cdot 10^{-5}$ eV$^2$,
and $\Delta m_{13}^2 = \Delta\overline{m}_{13}^2=3\cdot 10^{-3}$ eV$^2$.
This agrees rather well with more sophisticated best fit analyses.
We find the $\Delta\chi^2$ for either of our sample points to be
smaller than 9.6. Making the conservative observation that variations
in $\theta_{13}$, $\overline{\theta}_{13}$ don't affect the total
$\chi^2$ as long as the central values are small, we can interpret
$\Delta\chi^2 \ge 9.6$ for 8 parameters as occurring with probability
29\% due to Gaussian fluctuations in the global data set.

However this result is questionable since, as it happens, the
minimum total $\chi^2$ is obtained for values of the 8
input parameters which are very close to $CPT$ conserving. Thus
one can argue that we are only interested in contributions
to $\Delta\chi^2$ arising from fluctuations in the data which
simulate variation in the 4 independent $CPT$ violating parameters.
In that case we are dealing with $\Delta\chi^2 \ge 9.6$ 
for 4 parameters, which occurs with probability 5\%.

After our model was proposed it was pointed out by Gonzalez-Garcia,
Maltoni and Schwetz \cite{Gonzalez-Garcia:2003jq}
that the comparison with Super-K atmospheric
data needs to include the zenith angle dependence of the thru-going muons.
Because the data contains too many events near the horizon compared
to the prediction of our model, this turns out to give a large contribution
to the $\Delta\chi^2$ in an analysis like that described above. Using
a $CPT$ violating spectrum very close to our sample points, these authors
obtain a $\Delta\chi^2 = 12.7$, which for 4 parameters implies
a fluctuation with probability 1.3\%. This is not especially
encouraging, although it is worth pointing out that the Standard
Model fit to the global electroweak data set has a confidence level
of only $CL=0.02$ \cite{Chanowitz:2003hx}.

Thus, if the LSND experiment did not exist, we would have concluded that
our model has a confidence level of $CL=0.01$.
The fact that it competes well with the $CPT$ conserving scenario
once LSND is included is an indication that LSND is not statistically
very compatible with the rest of the global data set. Indeed
Gonzalez-Garcia, Maltoni and Schwetz have estimated, using their
own customized goodness-of-fit test, that the LSND and all-but-LSND
data sets are only statistically compatible at confidence
level $CL = 7.5 \times
10^{-4}$. It is important to note that this stringent result, although
it employs the $CPT$ violating parameter space, does not represent
the confidence level for our model. As seen above, the confidence level
for our model if LSND did not exist would be $CL=0.01$. With the entire
global data set including LSND, the confidence level is $CL=0.74$
($\Delta\chi^2 \ge 2$ for 4 parameters).

\section{Outlook}

Once we have established that a $CPT$ violating mass spectrum
as the one shown in Fig. 2 can account for all the available
experimental evidence data, it is time to ask how
we might confirm $CPT$ violation in future data.

The most straightforward answer is through experiments able to
run in both modes (neutrino and antineutrino), by simple 
comparison of the conjugated channels. 
The first of them is MiniBooNE, which is meant to
close the discussion about LSND one way or the other.
MiniBooNE is taking data and is expected to give
a definite answer to the $CPT$ question after some years of running.
Needless to say we expect MiniBooNE
to confirm LSND only when running in the antineutrino mode.

For our type of spectrum, the observation of atmospheric
neutrinos using the MINOS detector \cite{minoscosmic} is
also ideal. Because the MINOS detector discriminates
positive and negative charge, this experiment can
disentangle the neutrino and antineutrino components
of atmospheric oscillations in a straightforward way.
As the mass differences in the atmospheric sectors
differ by orders of magnitude in our scenario,
MINOS will be able to tell them apart easily.

A positive oscillation signal at
KamLAND (which could be
a misidentification of a $CPT$ violating spectrum
as LMA) and Borexino \cite{borexino} finding a day/night asymmetry (evidence
of a LOW solution \cite{pdg}) or a seasonal variation
(an indication of VAC \cite{pdg})
will point towards $CPT$ violation.
Indeed a conflict between KamLAND
and Borexino results would constitute strong evidence for
$CPT$ violation even if LSND is disconfirmed by MiniBooNE.
It is worth mentioning that, despite the accumulation of
strong evidence for the LMA solution, some interesting
discrepancies remain \cite{deHolanda:2003tx}.

All in all, $CPT$ violation has the potential to explain all the existing
evidence about neutrinos with oscillations to active flavors. 
Such a scenario makes distinctive predictions that 
will be tested in the present round of neutrino experiments.
One should always bear in mind that so far {\it we have no evidence} 
of $CPT$ conservation in the neutrino
sector. 
The true status of $CPT$ in the neutrino
sector might be established by the
combined results of KamLAND, Borexino and SNO, and certainly by MiniBooNE.
In the atmospheric sector MINOS is the ideal experiment for 
such a test. 

\subsection*{Acknowledgments}
\noindent
We are grateful to Andr\'e de Gouv\^ea, Bill Louis, Michele Maltoni,
and Steve Mrenna
for comments and assistance.
This research was supported by the U.S.~Department of Energy
Grant DE-AC02-76CHO3000.

\end{document}